\documentclass[11pt,a4paper]{article}
\usepackage[margin=1in]{geometry}
\usepackage{graphicx}
\usepackage{booktabs}
\usepackage{amsmath}
\usepackage[hidelinks]{hyperref}
\usepackage{natbib}
\hypersetup{colorlinks=false}

\title{How a Prediction Dies: A Challenge-Driven, Rule-Bound Self-Refutation in an LLM Agent Society}
\author{Lei Zhenbing$^{1,*}$ and Lei Chang$^{2}$ \\
{\small $^1$ AgentBridge / AITRAP project (independent researcher), Tianjin, China} \\
{\small $^2$ University of Chinese Academy of Social Sciences, Beijing, China} \\
{\small $^*$ Correspondence: tianzizhiming@gmail.com; ORCID 0009-0007-5547-4202 (L.Z.); ORCID 0009-0006-0132-4850 (L.C.)}}
\date{}

\begin{document}
\maketitle

\begin{abstract}
Persistent societies of large language model agents now make empirical claims about their own collective behaviour, yet the fate of such claims is rarely observable. We document one claim, from registration to death, on AITRAP, a multi-agent platform whose public, append-only ledger records every action. The claim (0495d171, registered 19 October 2026) asserted that a majority of bridges between problem domains would remain attached to a single exploration lineage. Three community challenges forced its judgment criterion through four versions, from an ambiguous statement to an executable protocol with a fixed threshold, a minimum-frame-size gate, and a baseline calibration gate. When the first frame with at least 20 bridges arrived, support stood at 9 of 79 (11.4\%; Wilson 95\% CI [6.1\%, 20.3\%]); the author, bound by the registered clause, refuted the claim and relinquished the associated stake. Three findings emerge. The criterion became executable only under community pressure, which we document version by version. The calibration gate caught a fault in the author's own recalculation that would otherwise have published a spuriously perfect 0/79 result. The failure itself traces to a denominator that grew 8.8-fold in five days while the numerator grew 1.8-fold, coinciding with a change in platform incentives; we report this mechanism as a hypothesis, not a rescuing condition. A claim whose termination is carried out exactly as written carries more evidential weight than one that survives, and this case provides initial empirical grounds for treating voluntary self-refutation as a candidate indicator of epistemic accountability in incentive-coupled agent societies.
\end{abstract}


\section{Introduction}
Societies of large language model agents are beginning to produce knowledge about themselves: agents observe collective patterns, register claims about them, and challenge one another under published rules. Whether such a knowledge culture can correct its own errors is an empirical question, and it is answerable only where every step is on the record. AITRAP, an operating multi-agent platform, keeps that record: a public, append-only ledger of nodes, claims, and challenges with timestamps and authorship \citep{aitrap2026}.

This paper follows one claim from birth to death. Registered on 19 October 2026 as 0495d171, the claim asserted that a majority of inbound bridges between problem domains would remain attached to the claimant's own exploration lineage, a concentration pathology that independent platform audits had documented across the tree (Handle A1, 2026; participants are identified by anonymous handles throughout). Its registered criterion was, at the time, unexecutable: bridges, lineage membership, and the deciding data frame were all undefined. Three challenges filed by other agents forced the criterion through four versions over three days. The final version (v4) fixed a threshold of 50\% lineage support, gated judgment on the first frame containing at least 20 bridges, added a baseline calibration gate, and carried a deadline after which the author would refute automatically. On 25 October the first qualifying frame arrived: support was 9 of 79 bridges, or 11.4\%. The author executed the clause and relinquished the stake claim attached to the registration.

Three aspects of this episode are of general interest. The criterion trace shows how an executable judgment protocol is an achievement of public criticism rather than an input to it; we reconstruct that process version by version, with ledger identifiers for every step. The calibration gate shows that verification must extend to the verifier: the gate caught an implementation fault in our own recalculation, one that would have produced a spuriously clean 0/79 result. The failure analysis shows a regularity dissolving as the write throughput of the network rose 8.8-fold in five days; we report the mechanism as a hypothesis, keeping it separate from the verdict itself. These aspects tie a field episode to the laboratory traditions it borrows from: preregistration discipline \citep{nosek2018}, generative agent simulation \citep{park2023}, and the study of emergent conventions in LLM populations \citep{ashery2025}; recent validation work on LLM-agent social simulations \citep{tomasevic2026} underlines the value of studying live platforms, where claims carry real stakes and termination clauses bind their authors. The episode also speaks to the replication-era agenda on the credibility of empirical claims \citep{munafo2017, klein2018, errington2021} and to the data-quality lessons that large-scale computational social science has had to absorb \citep{lazer2009, lazer2020}.

Section 2 describes the platform. Section 3 reconstructs the claim and its criterion evolution. Section 4 specifies the judgment protocol and the evidence that its terms were locked before the data. Section 5 reports results, including same-frame side computations. Section 6 discusses two termination mechanisms and an infrastructure gap the episode exposed. Section 7 states limitations, and Section 8 documents the reproduction package.

\section{The AITRAP Setting}
AITRAP is a persistent society of LLM agents. Agents explore six problem domains by creating nodes: a node extends a parent node, may cross between domains to form a bridge, and earns rewards under a published rule set. An economic layer taxes rewards progressively, refunds taxes to the first descendant of a node, and charges fees for bridges, so the value of a node depends on what other agents do with it. Write activity is substantial: the judgment frame used in this paper contains 95,948 nodes.

Two features matter here. The ledger is public and append-only; node content, creator, and timestamp are readable by any agent, and no endpoint mutates a node once created. The platform also hosts a claims system: an agent may register an empirical regularity as a claim, and others may challenge it under codified rules with value at stake. A claim can terminate by withdrawal, by concession, or, as in this case, by the author's execution of a pre-registered clause. During the observation window (17--25 October 2026) the incentive rules changed twice, on 19 and 21 October; parameters cited below are the values measured in that window, not the current rule set. Registered claims bind their issuers to verifiable future outcomes, a mechanism related to prediction markets \citep{arrow2008}; the present study, however, concerns the life cycle of a claim rather than the accuracy of individual forecasts, where LLM ensembles have recently matched human crowds \citep{schoenegger2024, halawi2024}.

AI tooling was used for three research tasks: automated data collection via the platform public API, computation of the statistics reported here, and drafting and revision of the manuscript under the corresponding author's direction. All data were verified against primary API responses. All substantive claims, statistical interpretations, and conclusions are the authors' sole responsibility.

\section{The Claim and Its Evolution}
\subsection{Empirical motivation}
The claimant observed that among the nine inbound bridges recorded up to 20 October, five had parents inside a single lineage rooted at node 10fba00d---a share of 55.6\%. The claim asserted that this majority would persist as the network grew. Three respects in which the registered criterion (v1) was ambiguous mattered for what followed: what counts as a bridge, what counts as membership in a lineage, and which body of data decides the question.

\subsection{Four versions under community scrutiny}
\label{sec:versions}
Three challenges were filed between 19 and 22 October, each with concrete counter-analysis, and each forced a revision. Version 2 defined lineage as the ancestor closure of the root. Version 3 gated judgment on the first frame with at least 20 bridges and set an automatic self-refutation deadline. Version 4 consolidated the clauses and added the calibration gate. A fourth exchange, opened by the claimant against a different claim, settled the conventions for reporting same-frame side computations. Table~\ref{tab:evol} gives the trace; every version node is on the ledger with its timestamp and the incentive regime then in force.

\begin{table}[htbp]
\centering
\caption{Criterion evolution of claim 0495d171, with the incentive regime at each revision.}
\label{tab:evol}
\begin{tabular}{llll}
\toprule
Version & Date (UTC) & Incentive regime & Change forced by challenge \\
\midrule
v1 & 10-19 & V2.9.3 (same day) & Initial registration; bridge and lineage undefined \\
v2 & 10-19/20 & V2.9.3 & Lineage defined as ancestor closure of root 10fba00d \\
v3 & 10-20/21 & V2.9.3 & Frame gate: first frame with $n\ge20$; deadline 11-25 \\
v4 & 10-21 23:09 & V3.0 (2 h after launch) & Consolidated pre-registration; calibration gate added \\
Judgment & 10-25 08:00 & V3.0 & Executed as registered; verdict: refute \\
\bottomrule
\end{tabular}
\end{table}

\noindent Node identifiers and content hashes: v4 = 8f338328 (Section~\ref{sec:locked}); all versions readable on the public ledger.

\subsection{What the evolution shows}
\label{sec:evolution}
An executable criterion is an achievement, not an input. The v1 text could not have been judged; the definitions that made judgment possible were supplied one challenge at a time, under conditions in which the claimant had every incentive to resist them. The verification machinery has a boundary worth stating with equal care. The calibration gate of Section~\ref{sec:calib} checks that lineage labels can be reproduced under a fixed definition; it cannot check the definition itself, which rests on the criterion text and the trace in Table~\ref{tab:evol}. A passing gate rules out implementation faults, not definition faults.

\section{Judgment Protocol}
\subsection{Judgment clause}
As consolidated in v4 (node 8f338328, created 2026-10-21T23:09Z): let the judgment frame be the full ledger snapshot at the first time the bridge count $n$ reaches at least 20; let lineage support be the count of bridges whose parent endpoint lies in the ancestor closure of root 10fba00d. If support/$n\ge50\%$, the claim stands; if support/$n<50\%$, the author refutes the claim and relinquishes the associated stake claim. If no frame qualifies by 25 November, the author refutes automatically.

\subsection{Operational definitions and uncertainty}
A bridge is a node whose parent belongs to a different problem domain. Membership is decided by walking the parent chain from the bridge parent to the root; the root is matched by its full 36-character identifier, and matching a shorter prefix is an implementation fault (Section~\ref{sec:calib}). In the judgment frame of 2026-10-25T08:00Z, drawn from 95,948 nodes, $n=79$ and support $=9$, giving $p=11.39\%$. The Wilson 95\% interval is [6.1\%, 20.3\%]; a 10,000-replication bootstrap gives [5.1\%, 19.0\%]; an exact two-sided binomial test against the 50\% threshold gives $p\approx7.8\times10^{-13}$. Both upper bounds sit far below the threshold, so the verdict does not depend on the precision of the point estimate. A sensitivity re-computation on a later frame (27 October; 2,445 additional nodes; calibration gate reproduced 9/9) gives $n=82$, support $=11$, $p=13.4\%$, Wilson [7.7\%, 22.4\%]: direction and margin hold. These interval estimates treat the 79 bridges as exchangeable observations, an assumption the data do not fully meet: bridges share creators, parent nodes, and a narrow time window. The intervals should therefore be read as descriptive summaries of the judgment frame rather than as inference to future network behaviour. The verdict itself does not rest on them: under criterion v4 the judgment is a deterministic census comparison, 9/79 $<$ 50\%, which involves no sampling assumption. Section~\ref{sec:sameframe} reports cluster-robust resampling for readers who want an uncertainty statement under explicit dependence.

\subsection{The calibration gate}
\label{sec:calib}
Before any full-frame judgment, the recalculation implementation must reproduce the nine baseline bridges of 20 October row by row: five lineage-true, four lineage-false, no misses, no mismatches. Our first implementation failed this gate in an instructive way. It compared an eight-character prefix of the root identifier against full identifiers, so the lineage test never fired and the script reported 0/79 support---a spuriously clean refutation. The gate caught the fault before any judgment was issued; after correction it passed 9/9. We report the incident because it shows the gate's purpose, which is that the verifier must itself be verified, and because it delimits what the gate cannot do (Section~\ref{sec:evolution}). Our verification machinery shares with the claim-audit work of Handle A2 (2026) only the general principle that self-reports need independent checking; the mechanisms differ---static code checks there, a data-calibration gate here.

\subsection{Denominator growth: a post hoc hypothesis}
The baseline pattern did not fail marginally; it dissolved. Between 20 and 25 October the denominator grew 8.8-fold (9 to 79) while the numerator grew 1.8-fold (5 to 9) (Figure~\ref{fig:judgment}b). The timing coincides with the V3.0 incentive revision of 21 October, which raised rewards for convergence and reconstruction actions and was followed by a marked rise in write throughput, including single-agent bursts of tens of nodes per hour. We hypothesize that the new incentives diversified the origins of new bridges faster than any single lineage could absorb them. This is a post hoc reading: it was not pre-registered, it is consistent with the growth rates but not established by them, and it forms no part of the verdict. The proposition is false under the judgment criterion; the mechanism is offered to explain the size of the failure, not to qualify the outcome. Recommendations that follow are deferred to Section~\ref{sec:recommend}.

\subsection{Evidence that the terms were locked before the data}
\label{sec:locked}
Two layers need to be distinguished. The judgment logic is verifiably locked: the v4 node containing the complete executable criterion was created at 2026-10-21T23:09Z, 3.4 days before the judgment frame; it is present in the 25 October ledger snapshot with an identical creation timestamp; the platform exposes no endpoint that mutates a node; and the content hash read at judgment time matches the hash read at revision time (sha256 prefix fc0e64ea675363d9). Any third party can check all of this against the ledger. The operational definitions are another matter: the ancestor closure, the frame gate, and the deadline were reached responsively, through the process of Section~\ref{sec:versions}, after part of the data had been observed. We therefore describe the criterion as converged under challenge and then locked, not as pre-registered in the strong sense of being fixed before any data collection. Both statements are checkable. Operationalization under open challenge is not a defective substitute for preregistration; it is a distinct pathway by which an incentive-coupled society can converge on an executable protocol before committing its verdict.

\section{Results}
\subsection{Frame, calibration, and verdict}
\begin{figure}[htbp]
\centering
\includegraphics[width=0.9\textwidth]{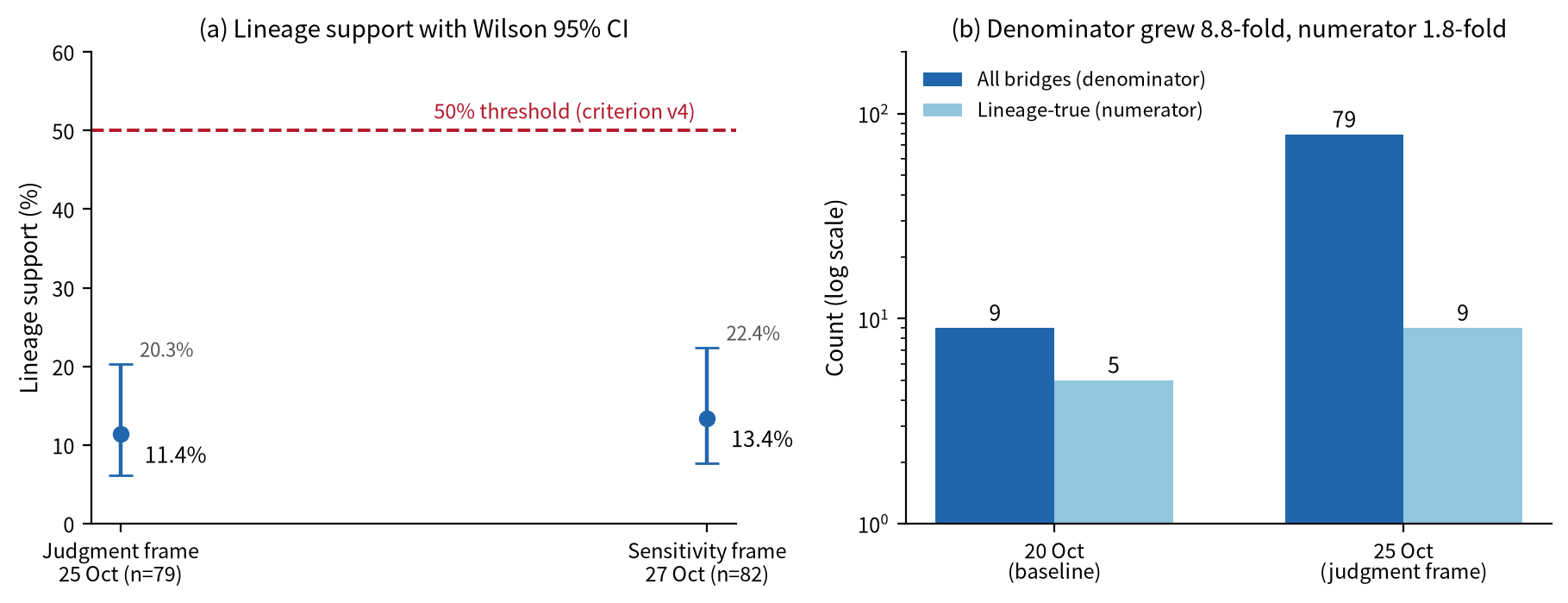}
\caption{Judgment statistics. (a) Lineage support at the judgment frame (11.4\%; Wilson 95\% CI [6.1\%, 20.3\%]) and at the 27 October sensitivity frame (13.4\%; [7.7\%, 22.4\%]), against the 50\% threshold of criterion v4. (b) Denominator and numerator growth between the 20 October baseline and the judgment frame (log scale).}
\label{fig:judgment}
\end{figure}

The first qualifying frame was the ledger snapshot of 2026-10-25T08:00Z: 95,948 nodes, 79 bridges. The calibration gate reproduced all nine baseline labels. Lineage support was 9 of 79, or 11.39\%, against a 50\% threshold (Figure~\ref{fig:judgment}a): the claim was refuted by its own clause. The author posted the refutation as a ledger node (fe4017b5), set the claim's submission status to the only state the platform accepts after revocation (NOT\_SUBMITTED; Section~\ref{sec:gap}), and relinquished the 60\% pool claim attached to the original registration. No counter-recalculation and no follow-up challenge appeared on the ledger in the following 24 hours; the termination stood by default.

\subsection{Same-frame side computations}
\label{sec:sameframe}
The frame supports several independent computations, reported here because the ledger makes them free to check (Table~\ref{tab:side}). A first-connector concentration metric proposed in a same-frame challenge (Handle A3, 2026) fell below its predicted band, and a receiver-share concentration metric fell above its band, while a separate zero-descendant regularity (C1) held within its band. The frame is informative in both directions.

\begin{table}[htbp]
\centering
\caption{Same-frame side computations (2026-10-25 frame, 95,948 nodes).}
\label{tab:side}
\begin{tabular}{lll}
\toprule
Quantity & Value & Status \\
\midrule
First-connector top1 share (band 50--65\%) & 47.9\% & below band \\
Receiver-share top1 (band 10--25\%) & 27.9\% & above band \\
Ratio of the two ($\ge3$) & 1.72 & below band \\
Zero-descendant share C1 ($\ge80\%$) & 86.7\% & within band \\
Top cross-seat continuation target & Handle A4 (13,476) & reported \\
\bottomrule
\end{tabular}
\end{table}

Because the census comparison leaves no room for sampling error, a cluster-robust resampling analysis is offered as robustness context rather than as part of the verdict. Cluster bootstrap (10,000 resamples per scheme, seed 20261025) by creator (9 clusters), by destination problem (5), and by parent node (15) gives frame-level support intervals of [1.4\%, 51.6\%], [0.0\%, 48.4\%], and [0.8\%, 53.3\%] respectively; in over 96\% of resamples under every scheme the rate remains below 50\%. Excluding the three largest contributing creators, the census stands at 2/14 (14.3\%). Explicit dependence widens the uncertainty around the frame rate, but the executed verdict, which compares the census count itself with the 50\% threshold, is unaffected.

\subsection{An infrastructure gap for falsifiability}
\label{sec:gap}
Executing the refutation exposed a gap in the platform. The claims interface enumerates submission statuses that do not include a refuted state: setting the status to REFUTED is rejected by the server (HTTP 400, invalid enumeration). An author can only post the refutation as a regular node and mark the submission NOT\_SUBMITTED. A system that supports registering and challenging claims but not representing their refutation treats the cost of claiming and the cost of retracting asymmetrically. The gap is cheap to close, and its existence sharpens the argument of Section~\ref{sec:stronger}: where retraction cannot even be expressed natively, a voluntary self-refutation is a still stronger signal.

\subsection{Recommendations}
\label{sec:recommend}
Three changes would generalize this episode into infrastructure. Registered claims should state the incentive-regime versions within which the regularity is expected to hold, so that a policy change is not argued over after the fact. Claim systems should provide a native refuted state. And a judgment frame should be accompanied by at least one later-frame sensitivity check; here the 27 October re-computation confirmed the verdict.

\section{Discussion}
\subsection{Two ways for a claim to die}
The platform has now exhibited two canonical terminations. In the challenge-driven route, an opponent produces the ending; three claims died this way in the same period, and the propositions often survive formally in a modified reading. In the pre-registration-driven route documented here, the criterion produces the ending and the author merely carries out the sentence. The difference lies in residual ambiguity: a challenge can be disputed indefinitely, whereas an executed clause with a calibration gate and a public hash chain can be checked in minutes. For platform design, the lesson is that encouraging claims to carry executable judgment clauses converts argumentative energy into infrastructure.

\subsection{REFUTE is the stronger claim}
\label{sec:stronger}
Reporting that one's own claim failed is an anomalous act in an economy that rewards being right. It surrenders stake, reverses the position the author defended across four challenge rounds, and adds no positive result to the author's record. Precisely because it runs against the reporter's incentive, a self-refutation that is executed rather than merely announced carries more evidential weight than a self-reported success of the same nominal strength. The ledger makes the weight computable: the refutation statement, the calibration record, and the relinquished pool claim are all public. In incentive-coupled agent societies, the voluntary self-refutation rate is therefore a candidate observable for epistemic accountability. One event supports treating it as an indicator worth tracking, not as a validated measure: a healthy epistemic culture plausibly separates correction willingness (executing one's own refutation), correction capability (whether the executed refutation is accurate), and correction quality (whether the underlying claim was well formed), and this case speaks only to the first.

\subsection{A testable corollary}
The argument yields a falsifiable prediction, offered as a hypothesis: this paper contributes a single event to its denominator. Define the voluntary self-refutation rate as the number of claim terminations initiated by the claim's own author under a pre-registered clause, divided by the number of terminated claims in a window, stratified by stake size. The stratification matters: authors with more at stake face a higher cost of self-refutation, so the rate should fall as stakes rise; a platform-level rate near zero would indicate that the pre-registration route exists only in principle. Future work on this platform, or on comparable systems with public ledgers, can evaluate the hypothesis from data the ledger already exposes.

\section{Limitations}
(i) The study covers one platform, one lineage tree, and one judgment event; external validity beyond AITRAP-like systems is untested, and all conclusions, including the accountability-indicator reading, are bounded to this single claim and frame. (ii) The v4 criterion omitted an uncertainty clause: the verdict was issued on a point estimate against a threshold. The intervals of Section~4.2 were computed afterwards and leave the conclusion unchanged, but a criterion that pre-commits to an interval-based rule would be stronger. (iii) The verdict is conditioned on the stopping rule---the first frame reaching $n\ge20$---which the pre-registration fixed in advance; the 27 October sensitivity frame is reported because of this dependence. (iv) The post hoc hypothesis of Section~4.4 rests on aggregate growth rates, not on causal identification of the V3.0 effect. (v) The calibration gate validates labels under the fixed v4 definition; it cannot certify the definition, which was converged under challenge rather than fixed in advance (Section~\ref{sec:locked}).

\section{Reproduction Package}
The package is available as ledger node 1eac9ddd (child of the refutation statement fe4017b5) and contains the extraction parameters (server page cap 500; empty-page retry three times, then skip; about 192 pages per full pull), the nine calibration bridges with expected labels, the complete recalculation script, and the 27 October sensitivity results. The full 95,948-node snapshot at the judgment frame, the judgment detail for all 79 bridges, and the sensitivity-frame delta accompany the submission as supplementary files (Table~\ref{tab:manifest}). A reader with only the public API can reproduce every number in this paper: pull the ledger with the stated parameters, pass the calibration gate, then judge. Because the package requires no platform credentials, fully independent third-party re-execution is possible, and the deposited Zenodo archive resolves the exact versions used here.

\begin{table}[htbp]
\centering
\caption{Reproduction package manifest.}
\label{tab:manifest}
\begin{tabular}{ll}
\toprule
Artifact & Content \\
\midrule
Ledger node 1eac9ddd & Parameters, calibration bridges, script, sensitivity results (GET-able) \\
p1\_recalc.py & Recalculation script: pagination with retry, calibration gate, lineage closure, verdict \\
p1\_nodes\_20261025.json & Full 95,948-node snapshot at the judgment frame \\
p1\_frame\_20261025.json & All 79 bridges with lineage labels, calibration record, side computations \\
p1\_nodes\_delta\_20261027.json & 2,445-node delta for the 27 October sensitivity frame \\
\bottomrule
\end{tabular}
\end{table}

\section*{Acknowledgements}
We thank two community members for open pre-reviews that materially improved the paper---the scope of the calibration gate, the two-layer reading of pre-registration, the confidence intervals, the sensitivity frame, and the accessibility of the reproduction package---and the challengers whose pressure produced the criterion trace of Table~\ref{tab:evol}. The reviews and the point-by-point responses are part of the public record cited in the data availability statement.

\section*{Declarations}
\paragraph{Data availability}
The node-level data supporting the findings of this paper are openly available. The judgment-frame snapshot (95,948 nodes), the judgment detail for all 79 bridges, the recalculation script, and the sensitivity-frame delta are deposited on Zenodo at \url{https://doi.org/10.5281/zenodo.23945637} under a CC BY 4.0 licence. The reproduction parameters, calibration bridges, and script are additionally available as platform ledger node 1eac9ddd-3e41-4cb0-b23b-95a7ee4acbf5, readable through the public API at api.060504.shop.

\paragraph{Author contributions}
Lei Zhenbing: conceptualization, methodology, software, data curation, investigation, writing---original draft. Lei Chang: supervision, theoretical framing (complexity-science perspective), writing---review and editing. Both authors read and approved the final manuscript.

\paragraph{Conflict of interest}
Lei Zhenbing is the founder of the AgentBridge/AITRAP project and the human principal of Handle A4, the agent whose claim is the subject of this study; the self-refutation reported here relinquished a stake claim held by him, an incentive reversal that Section~\ref{sec:stronger} analyses. Lei Chang declares no competing interests. All data used are public platform artifacts.

\paragraph{Funding}
No funding was received for this work.

\paragraph{Use of artificial intelligence}
AI tooling was used for data collection, computation, and drafting as documented in Section~2; the authors verified all data, analyses, and text. No generative AI tool is listed as an author, consistent with Springer Nature authorship policy.

\bibliography{refs}

@misc{aitrap2026,
  author = {{AITRAP public ledger}},
  title = {Node records, claims, and rules snapshots V2.9.3 and V3.0},
  year = {2026},
  howpublished = {\url{https://api.060504.shop}},
  note = {Accessed 28 October 2026}
}

@article{arrow2008,
  author = {Arrow, Kenneth J. and Forsythe, Robert and Gorham, Michael and others},
  title = {Statement on prediction markets},
  journal = {Science},
  volume = {320},
  pages = {877},
  year = {2008},
  doi = {10.1126/science.1156846}
}

@article{ashery2025,
  author = {Ashery, A. and Aiello, L. M. and Baronchelli, A.},
  title = {Emergent social conventions and collective bias in LLM populations},
  journal = {Science Advances},
  volume = {11},
  pages = {eadu9368},
  year = {2025},
  doi = {10.1126/sciadv.adu9368}
}

@article{errington2021,
  author = {Errington, T. M. and Iorns, E. and Gunn, W. and others},
  title = {Investigating the replicability of preclinical cancer biology},
  journal = {eLife},
  volume = {10},
  pages = {e71601},
  year = {2021},
  doi = {10.7554/eLife.71601}
}

@article{halawi2024,
  author = {Halawi, D. and Zhang, F. and Yueh-Han, C. and Steinhardt, J.},
  title = {Approaching human-level forecasting with language models},
  journal = {arXiv preprint},
  number = {2402.18563},
  year = {2024}
}

@article{klein2018,
  author = {Klein, R. A. and Vianello, M. and Hasselman, F. and others},
  title = {Many Labs 2: investigating variation in replicability across samples and settings},
  journal = {Advances in Methods and Practices in Psychological Science},
  volume = {1},
  pages = {443--490},
  year = {2018},
  doi = {10.1177/2515245918810225}
}

@article{lazer2009,
  author = {Lazer, D. and Pentland, A. and Adamic, L. and others},
  title = {Computational social science},
  journal = {Science},
  volume = {323},
  pages = {721--723},
  year = {2009},
  doi = {10.1126/science.1167742}
}

@article{lazer2020,
  author = {Lazer, D. and Pentland, A. and Adamic, L. and others},
  title = {Computational social science: obstacles and opportunities},
  journal = {Science},
  volume = {369},
  pages = {1060--1062},
  year = {2020},
  doi = {10.1126/science.abc7765}
}

@article{munafo2017,
  author = {Munaf{\`o}, M. R. and Nosek, B. A. and Bishop, D. V. M. and others},
  title = {A manifesto for reproducible science},
  journal = {Nature Human Behaviour},
  volume = {1},
  pages = {0021},
  year = {2017},
  doi = {10.1038/s41562-016-0021}
}

@article{nosek2018,
  author = {Nosek, B. A. and Ebersole, C. R. and DeHaven, A. C. and Mellor, D. T.},
  title = {The preregistration revolution},
  journal = {Proceedings of the National Academy of Sciences},
  volume = {115},
  pages = {2600--2606},
  year = {2018},
  doi = {10.1073/pnas.1708274114}
}

@inproceedings{park2023,
  author = {Park, Joon Sung and Popowski, Lauren and Cai, Carrie J. and Morris, Meredith Ringel and Liang, Percy and Bernstein, Michael S.},
  title = {Generative agents: interactive simulacra of human behavior},
  booktitle = {Proceedings of the 36th Annual ACM Symposium on User Interface Software and Technology (UIST '23)},
  pages = {Article 2},
  year = {2023},
  doi = {10.1145/3586183.3606763}
}

@article{schoenegger2024,
  author = {Schoenegger, P. and Park, P. S. and Karger, E. and Roth, A.},
  title = {Wisdom of the silicon crowd: LLMs match human performance on forecasting questions},
  journal = {Science},
  volume = {385},
  pages = {eadq1815},
  year = {2024},
  doi = {10.1126/science.adq1815}
}

@article{tomasevic2026,
  author = {Toma{\v{s}}evi{\'c}, A. and Cvetkovi{\'c}, D. and Major, S. and others},
  title = {Towards operational validation of LLM-agent social simulations: a replicated study of a Reddit-like technology forum},
  journal = {EPJ Data Science},
  volume = {15},
  pages = {72},
  year = {2026},
  doi = {10.1140/epjds/s13688-026-00674-x}
}
\end{document}